\documentstyle[12pt]{article}
\input psfig
\newcommand{\bea}{\begin{eqnarray}}
\newcommand{\eea}{\end{eqnarray}}
\newcommand{\beq}{\begin{equation}}
\newcommand{\eeq}{\end{equation}}

\def\msbar{\ifmmode{\overline{\rm MS}} \else{$\overline{\rm MS}$} \fi}
\def\drbar{\ifmmode{\overline{\rm DR}} \else{$\overline{\rm DR}$} \fi}
\def\st{\ifmmode{{\tilde{t}}} \else{${\tilde{t}}$} \fi}
\def\sb{\ifmmode{{\tilde{b}}} \else{${\tilde{b}}$} \fi}
\def\sq{\ifmmode{{\tilde{q}}} \else{${\tilde{q}}$} \fi}
\def\sg{\ifmmode{{\tilde{g}}} \else{${\tilde{g}}$} \fi}

\newcommand{\smallfrac}{\mbox{\small $\frac{1}{2}$}}
\newcommand{\al}{\alpha}
\newcommand{\be}{\beta}
\newcommand{\si}{{\mbox{s}}}
\newcommand{\co}{{\mbox{c}}}

\begin{document}
\pagestyle{empty}
\vspace*{-2cm}
\begin{flushright}
UWThPh-1997-03\\
HEPHY-PUB 664/97\\
TGU-19\\
ITP-SU-97/01\\
TU-517\\
hep-ph/9701398\\
\end{flushright}

\vspace{1cm}
\begin{center}
\begin{Large} {\bf
QCD corrections to\\ 
Higgs boson decays into squarks in\\
the Minimal Supersymmetric Standard Model}\\
\end{Large}
\end{center}
\vspace{10mm}

\begin{center}
\large A.~Bartl,$^1$ H.~Eberl,$^2$ K.~Hidaka,$^3$ T.~Kon,$^4$\\
W.~Majerotto$^2$ and Y.~Yamada$^5$
\end{center}
\vspace{0mm}

\begin{center}
\begin{tabular}{l}
$^1${\it Institut f\"ur Theoretische Physik, Universit\"at Wien, A-1090
Vienna, Austria}\\
$^2${\it Institut f\"ur Hochenergiephysik der \"Osterreichischen Akademie
der Wissenschaften,}\\
{\it A-1050 Vienna, Austria}\\
$^3${\it Department of Physics, Tokyo Gakugei University, Koganei,
Tokyo 184, Japan}\\
$^4${\it Faculty of Engineering, Seikei University, Musashino, Tokyo 180,
Japan}\\
$^5${\it Department of Physics, Tohoku University, Sendai 980--77, Japan}
\end{tabular}
\end{center}

\vfill

\begin{abstract}
\begin{small}
\baselineskip=21pt
We calculate the supersymmetric ${\cal O}(\alpha_s)$ QCD corrections to the
widths of the Higgs boson decays $H^+ \to \st \bar\sb$ and
$H^0, A^0 \to \st \bar\st, \sb \bar\sb$ in the on--shell scheme within  
the Minimal Supersymmetric Standard
Model. We find that
the QCD corrections are significant, but that the squark pair decay modes are
still domi\-nant  
in a wide parameter
region.
\end{small}
\end{abstract}

\newpage
\pagestyle{plain}
\setcounter{page}{2}

\section{Introduction}

It is well known that
the Minimal
Supersymmetric Standard Model (MSSM) \cite{Haber} predicts the
existence of five physical Higgs bosons $h^0, H^0, A^0$, and $H^\pm$
\cite{Gunion1,Gunion2}. 
In order to facilitate experimental Higgs boson searches it is
necessary to perform a thorough theoretical study of their decay
branching ratios 
\cite{Kunszt}. Their decays into supersymmetric
(SUSY) particles can be very important if they are kinematically
allowed \cite{Djouadi1,Bartl1,Djouadi2,Bartl2}. The decays into the
3rd--generation squarks $\st$ and $\sb$ can play a special
r\^ole because they can be much lighter than the other squarks and the decays 
can be strongly enhanced due to their large Yukawa couplings and their
large $\sq_L - \sq_R$ mixings \cite{Bartl1,Djouadi2,Bartl2}.
The tree--level results of  
Refs.~\cite{Bartl1,Bartl2} show that the decay modes
$H^+ \to \st \bar\sb$ and $H^0, A^0 \to \st \bar\st, \sb \bar\sb$ can
be dominant in a large region of the parameter space of the MSSM,
and that this could have an important impact on searches for $H^+, H^0$,
and $A^0$ at future colliders.\\
The SUSY QCD corrections to the decays $H^+ \to t \bar b$ 
and $H^0, A^0 \to t \bar t, b \bar b$ can
be large \cite{Jimenez}. This suggests that the QCD corrections to
$H^+ \to \st \bar\sb$ and $H^0, A^0 \to \st \bar\st, \sb \bar\sb$ could also be
large. Therefore it is very important to examine whether the results of
Refs.~\cite{Bartl1,Bartl2} remain valid after including the QCD corrections.
In Ref.~\cite{Bartl3} it was shown that the QCD corrections to 
$H^+ \to \st \bar\sb$ can be significant in the \drbar renormali\-zation  
scheme,
but that they do not invalidate the result in Ref.~\cite{Bartl1} on the
dominance of the $H^+ \to \st \bar\sb$ mode in a large parameter region.\\

In  the present paper we extend our study to the decays of the charged
and neutral Higgs bosons.
We use the
on--shell scheme which is more appropriate for the discussion of 
physical observables.
We calculate the complete ${\cal O}(\alpha_s)$ QCD corrections to the widths of 
the decays $H^+ \to \st \bar\sb$ and $H^0, A^0 \to \st \bar\st, \sb \bar\sb$
within the MSSM including all quark mass terms and
$\sq_L - \sq_R$ mixings. The main complication here is that the
$\sq_L - \sq_R$ mixing angles are renormalized by the SUSY--QCD corrections.
This problem was first solved in Ref.~\cite{Eberl} in the treatment
of $e^+ e^- \to \sq \bar\sq$, where a suitable renormalization condition for 
the squark--mixing angle was found. The method was also applied in
\cite{Kraml,Djouadi3,Oakes,Beenakker} to 
$\sq_i \to q \tilde{\chi}_j^{0,\pm}$ and
$t \to \st_1 \tilde{\chi}_1^0$. In the present study 
we use the renormalization prescription 
as given in \cite{Eberl}.
Furthermore, we point out that
special attention must be
paid to the soft SUSY--breaking parameter $M_{\tilde Q}$, 
which enters the stop and sbottom mass matrices: in the on--shell scheme,
the renormalized $M_{\tilde Q}$ in the
stop sector is different from that in the sbottom sector.\\
We find that the QCD corrections to the squark pair widths are significant,
but that the squark pair modes
($H^+ \to \st \bar\sb$ and $H^0, A^0 \to \st \bar\st, \sb \bar\sb$) are
still dominant in a wide parameter range. 


\section{Tree level result}
We first review the tree level results \cite{Bartl1,Bartl2}. 
The squark mass matrix in
the basis ($\sq_L$, $\sq_R$),
with $\sq=\st$ or $\sb$, is given by \cite{Gunion1,Gunion2} 
\beq \label{1}
\left( \begin{array}{cc}m_{LL}^2 & m_{LR}^2 \\ m_{RL}^2 & m_{RR}^2
\end{array} \right)=
(R^{\sq})^{\dagger}\left( \begin{array}{cc}m_{\sq_1}^2 & 0 \\ 0 &
m_{\sq_2}^2 \end{array}
\right)R^{\sq}, 
\eeq
where
\bea
m_{LL}^2 &=& M_{\tilde{Q}}^2+m_q^2+m_Z^2\cos 2\beta 
(I^{3L}_q-e_q\sin^2\theta_W), \\
m_{RR}^2 &=& M_{\{\tilde{U},
\tilde{D}\}}^2+m_q^2+m_Z^2\cos 2\beta e_q\sin^2
\theta_W, \\
m_{LR}^2=m_{RL}^2 &=& \left\{ \begin{array}{ll}
m_t(A_t-\mu\cot\beta) & (\sq=\st) \\
m_b(A_b-\mu\tan\beta) & (\sq=\sb) \end{array} \right. ,\label{4} \eea
and
\beq \label{5}
R^{\sq}_{i\alpha}=\left(
\begin{array}{cc}\cos\theta_\sq & \sin\theta_\sq \\
-\sin\theta_\sq & \cos\theta_\sq \end{array}\right) . \eeq
The mass eigenstates $\sq_i(i=1,2)$ (with $m_{\sq_1}<m_{\sq_2}$) are
related to the
SU(2)$_L$ eigenstates $\sq_{\alpha}(\alpha=L,R)$ by 
$\sq_i=R^{\sq}_{i\alpha}\sq_{\alpha}$. 

\noindent The tree--level decay width 
of $H^k \rightarrow \sq_i \bar{\sq}_j$ is then
given by (see Fig.~1a)
\beq \label{6}
\Gamma^{tree}(H^k\rightarrow\sq_i\bar{\sq}_j) =\frac{N_C\kappa}{16\pi
m_{H^k}^3}|G_{ijk}^\sq|^2\,. \eeq
For $k = 1, 2, 3$ $H^k$ denotes the neutral Higgs bosons (i. e.
$H^1 \equiv h^0$,
$H^2 \equiv H^0$,
$H^3 \equiv A^0$)
and $\sq = \st, \sb$. 
For $k = 4$ one has $H^4 \equiv H^+$ and  $\sq_i \equiv \st_i$,
$\sq_j \equiv \sb_j$, and the upper index $\sq$ in $G_{ijk}^\sq$ is omitted.
$\kappa=\kappa(m_{H^k}^2, m_{\sq_i}^2, m_{\sq_j}^2 )$, $\kappa(x,y,z)\equiv
((x-y-z)^2-4yz)^{1/2}$, $N_C=3$,
and the $H^k \sq_i^* \sq_j$ couplings \cite{Gunion1,Gunion2} are given by  
\beq \label{7}
\begin{array}{l}
G_{ij2}^\st \equiv 
G(H^0 \st_i^* \st_j) =
\left(R^\st\, G_{LR}^\st\, (R^{\st})^T 
\right)_{ij} = \\
{\normalsize -g R^{\st}
\left( \begin{array}{cc}
\frac{m_Z}{c_W}(I_t^{3 L} - 
e_t s_W^2) \co_{\al + \be} + \frac{m_t^2}{m_W\si_\be}
\si_\al & \frac{m_t}{2 m_W \si_\be} (A_t \si_\al - \mu \co_\al)\\
\frac{m_t}{2 m_W \si_\be} (A_t \si_\al - \mu \co_\al) &
\frac{m_Z}{c_W} e_t s_W^2 \co_{\al + \be} + \frac{m_t^2}{m_W \si_\be} \si_\al
\end{array}
\right)(R^{\st})^T}\, , \nonumber 
\end{array}
\eeq
\beq \label{8}
\begin{array}{l}
G_{ij2}^\sb \equiv 
G(H^0 \sb_i^* \sb_j) =
\left(R^\sb\, G_{LR}^\sb\, (R^{\sb})^T 
\right)_{ij} = \\
{\normalsize
 -g R^{\sb}
\left( \begin{array}{cc}
\frac{m_Z}{c_W}(I_b^{3 L} - e_b s_W^2) \co_{\al + \be} + \frac{m_b^2}
{m_W\co_\be}
\co_\al & \frac{m_b}{2 m_W \co_\be} (A_b \co_\al - \mu \si_\al)\\
\frac{m_b}{2 m_W \co_\be} (A_b \co_\al - \mu \si_\al) &
\frac{m_Z}{c_W} e_b s_W^2 \co_{\al + \be} + \frac{m_b^2}{m_W \co_\be} \co_\al
\end{array}
\right)(R^{\sb})^T}\, , \nonumber 
\end{array}
\eeq
\bea 
& G_{ij1}^\sq \equiv G(h^0 \sq_i^* \sq_j) = 
\{ G(H^0 \sq_i^* \sq_j) \mbox{ with } \al \to
\al + \smallfrac \pi\,, 
\nonumber\\
&\mbox{(i. e.}\ \sin\al \equiv \si_\al \to \co_\al\,,\ 
\cos\al \equiv \co_\al \to -\si_\al\,,\ \mbox{and}
\cos(\al+\be) \equiv \co_{\al+\be} \to -\si_{\al+\be})\}\,, \label{9}
\eea
\bea \label{10}
G_{ij3}^\st &=& G(A^0 \st_i^* \st_j) =
\frac{i g}{2 m_W} \left( \begin{array}{cc} 
0 & m_t (A_t \cot\be + \mu)\\
-m_t (A_t \cot\be + \mu) & 0
\end{array}
\right)\, , \nonumber\\
\eea
\bea \label{11}
G_{ij3}^\sb &=& G(A^0 \sb_i^* \sb_j) =
\frac{i g}{2 m_W} \left( \begin{array}{cc} 
0 & m_b (A_b \tan\be + \mu)\\
-m_b (A_b \tan\be + \mu) & 0
\end{array}
\right)\, , 
\eea

\beq  \label{H+tree}
\begin{array}{l}
G_{ij4} \equiv G(H^+ \st_i^* \sb_j) =\\ 
\frac{g}{\sqrt{2}m_W}
R^{\st}\left( \begin{array}{cc}
m_b^2\tan\beta+m_t^2\cot\beta-m_W^2\sin 2\beta  &
m_b(A_b\tan\beta+\mu) \\
m_t(A_t\cot\beta+\mu) &
2m_tm_b/\sin 2\beta \end{array} \right)(R^{\sb})^T \, .
\end{array}
\eeq

\noindent Here $g$ is the SU(2) coupling,
$\alpha$ is the mixing angle in the CP even neutral 
Higgs boson sector, $c_W \equiv \cos\theta_W$ and
$s_W \equiv \sin\theta_W$.

\section{QCD corrections}
The ${\cal O}(\alpha_s)$ QCD virtual corrections to
$H^k\rightarrow\sq_i\bar{\sq}_j$ stem from the
diagrams of Fig.~1b (vertex corrections) and Fig.~1c (wave--function
corrections). All parameters of the QCD interacting
particles, appearing in the tree--level mass matrix of eqs.~(\ref{1}) --
(\ref{5}) and the tree--level couplings of eqs.~(\ref{7}) -
(\ref{H+tree}), have to be renormalized.
These are the 
soft--SUSY--breaking squark masses $M_{\tilde Q,\tilde U, \tilde D}$,
the quark masses $m_{t,b}$,
the trilinear couplings 
$A_{t,b}$, the squark masses $m_{\sq_{1,2}}$, and the 
mixing angles $\theta_{\st,\sb}$ ($\alpha$, $\beta$, and $\mu$ are 
of course not renormalized by QCD).
In this paper we use the on--shell renormalization scheme.\\

\noindent The one--loop corrected decay amplitudes 
$G_{ijk}^{\sq\,\rm corr}$ can be expressed as
\beq \label{13}
G_{ijk}^{\sq\, \rm corr}=G_{ijk}^\sq
+\delta G_{ijk}^{\sq\, (v)}+\delta G_{ijk}^{\sq\, (w)}+
\delta G_{ijk}^{\sq\, (0)}, \eeq
where $G_{ijk}^\sq$ are defined by eqs.~(\ref{7}) -- (\ref{H+tree}) 
in terms of the on--shell parameters, and
$\delta G_{ijk}^{\sq\,  (v)}$ and
$\delta G_{ijk}^{\sq\,  (w)}$ 
are the vertex and squark wave--function corrections,
respectively. $\delta G_{ijk}^{\sq\, (0)}$ denotes the counterterm caused by
the on--shell renormalization.
They get contributions from the gluon, gluino and squark exchange.
(Again the upper index $\sq$ has to be omitted for $k = 4$.)

\noindent The gluon exchange contributions to the vertex corrections are
(see Fig.~1b)
\bea
\delta G_{ijk}^{\sq\, (v, g)}&=& \frac{\alpha_s C_F}{4\pi} G_{ijk}^\sq
\left[ B_0(m_{\sq_i}^2,
0, m_{\sq_i}^2)+B_0
(m_{\sq_j}^2, 0, m_{\sq_j}^2) -B_0(m_{H^k}^2, m_{\sq_i}^2, m_{\sq_j}^2)
\right. \nonumber \\
&& -2(m_{H^k}^2-m_{\sq_i}^2-m_{\sq_j}^2)
C_0(\lambda^2, m_{\sq_i}^2, m_{\sq_j}^2)\left.\right] \, .
\eea

\noindent The gluino exchange contributions to them are (see Fig.~1b)
\bea \nonumber
\delta G_{ijk}^{\sq\, (v, \sg)}&=& \frac{\alpha_s C_F}{2\pi} s_k^q \left[\right.
2 \delta_{ij} m_q B_0(m_{H^k}^2, m_q^2, m_q^2) \label{14} \\
&& + \{ \delta_{ij} m_q + S_{ij}^\sq m_\sg \}
(B_0(m_{\sq_i}^2, m_{\sg}^2, m_q^2) +
B_0(m_{\sq_j}^2, m_{\sg}^2, m_q^2))+\{ (4 m_q^2 -\nonumber\\ 
&& m_{H^k}^2) S_{ij}^\sq m_{\sg}+
( 2 m_q^2 + 2 m_{\sg}^2 - m_{\sq_i}^2 - m_{\sq_j}^2) \delta_{ij} m_q \}
C_0( m_{\sg}^2, m_q^2, m_q^2) \left.\right]\, ,
\eea
\noindent for $k = 1,2$, 
\bea \label{15} \nonumber
\delta G_{ij3}^{\sq\, (v, \sg)}&=& -\frac{\alpha_s C_F}{2\pi} a^q \left[\right.
A_{ij}^\sq m_q (B_0(m_{\sq_i}^2, m_{\sg}^2, m_q^2) -
B_0(m_{\sq_j}^2, m_{\sg}^2, m_q^2)) \\ 
&& + \epsilon_{ij} m_\sg (B_0(m_{\sq_i}^2, m_{\sg}^2, m_q^2) +
B_0(m_{\sq_j}^2, m_{\sg}^2, m_q^2)) \nonumber\\ 
&& +\{ (m_{\sq_i}^2 - m_{\sq_j}^2)
A_{ij}^\sq m_q - m_{H^k}^2 \epsilon_{ij} m_{\sg}
\} C_0 ( m_{\sg}^2, m_q^2, m_q^2) \left.\right]\, ,
\eea
and
\bea \label{15a}
\delta G_{ij4}^{(v, \sg)}&=& \frac{\alpha_s C_F}{2\pi}\left[\right.
\{ (\alpha_{LL})_{ij}(m_ty_2+m_by_1)+(\alpha_{RR})_{ij}(m_ty_1+m_by_2)\}
B_0(m_H^2, m_b^2, m_t^2) \nonumber\\
&& +\{ (\alpha_{LL})_{ij}m_ty_2+(\alpha_{LR})_{ij}m_{\sg}y_1
+(\alpha_{RL})_{ij}m_{\sg}y_2+(\alpha_{RR})_{ij}m_ty_1\}
B_0(m_{\st_i}^2, m_{\sg}^2, m_t^2) \nonumber\\
&& +\{ (\alpha_{LL})_{ij}m_by_1+(\alpha_{LR})_{ij}m_{\sg}y_1
+(\alpha_{RL})_{ij}m_{\sg}y_2+(\alpha_{RR})_{ij}m_by_2\}
B_0(m_{\sb_j}^2, m_{\sg}^2, m_b^2) \nonumber\\
&& +\{ (m_t^2+m_b^2-m_H^2)m_{\sg}
        ((\alpha_{LR})_{ij}y_1+(\alpha_{RL})_{ij}y_2) \nonumber\\
&& +(m_b^2+m_{\sg}^2-m_{\sb_j}^2)m_t
        ((\alpha_{LL})_{ij}y_2+(\alpha_{RR})_{ij}y_1) \nonumber\\
&& +(m_t^2+m_{\sg}^2-m_{\st_i}^2)m_b
        ((\alpha_{LL})_{ij}y_1+(\alpha_{RR})_{ij}y_2) \nonumber\\
&& \left. +2m_{\sg}m_tm_b((\alpha_{LR})_{ij}y_2+(\alpha_{RL})_{ij}y_1) \}
C_0(m_{\sg}^2, m_t^2, m_b^2) \right]\, .
\eea

\noindent The vertex corrections due to the four--squark interaction are
(see Fig.~1b)
\bea
\delta G_{ijk}^{\sq\, (v, \sq)}&=& -\frac{\alpha_s C_F}{4\pi} 
\sum_{n,l} B_0(m_{H^k}^2, m_{\sq_n}^2,
m_{\sq_l}^2)G_{nlk}^\sq A^{\sq}_{in} A^{\sq}_{lj} \, , \quad
(k = 1,2,3),
\eea
\bea
\delta G_{ij4}^{(v, \sq)}&=& -\frac{\alpha_s C_F}{4\pi} 
\sum_{n,l} B_0(m_{H^+}^2, m_{\st_n}^2,
m_{\sb_l}^2)G_{nl4} A^{\st}_{in} A^{\sb}_{lj} \, .
\eea

\noindent
$s^q_k$ $(k = 1,2)$, $a^q$, $y_1$, and $y_2$ in eqs.~(\ref{14}), (\ref{15}),
and (\ref{15a}) are the Yukawa couplings \cite{Gunion1}:
\beq
{\cal L} = s_1^q {h^0} \bar{q} q + s_2^q {H^0} \bar{q} q +
a_q {A^0} \bar{q} \gamma^5 q + H^+ \bar t (y_1 P_R + y_2 P_L) b\, ,
\eeq
with
\beq
\begin{array}{lcl}
s_1^t = -g\frac{m_t \cos\al}{2 m_W \sin\be}\, , &&
s_1^b = g\frac{m_b \sin\al}{2 m_W \cos\be}\, ,\\[2mm]
s_2^t = -g\frac{m_t \sin\al}{2 m_W \sin\be}\, , &&
s_2^b = -g\frac{m_b \cos\al}{2 m_W \cos\be}\, ,\\[2mm]
a^t = i g\frac{m_t \cot\be}{2 m_W}\,, &&
a^b = i g\frac{m_b \tan\be}{2 m_W}\, , \\[2mm]
y_1=\frac{g}{\sqrt{2}m_W}m_b\tan\beta=h_b\sin\beta &\mbox{and}& 
y_2=\frac{g}{\sqrt{2}m_W}m_t\cot\beta=h_t\cos\beta\, .
\end{array}
\eeq
 
\noindent $C_F=4/3$, $\delta_{ij}$ is the unit matrix, $\epsilon_{ij}$ is 
totally antisymmetric with $\epsilon_{12} = 1$,
\beq
A^{\sq}=\left( \begin{array}{cc} \cos 2\theta_\sq & -\sin 2\theta_\sq \\
-\sin 2\theta_\sq & -\cos 2\theta_\sq
\end{array} \right)\, , \, 
S^{\sq}=\left( \begin{array}{cc} -\sin 2\theta_\sq & -\cos 2\theta_\sq \\
-\cos 2\theta_\sq & \sin 2\theta_\sq
\end{array} \right) = \epsilon_{ik} A_{kj}^{\sq}\, ,
\eeq
\beq
(\alpha_{\sigma \rho})_{ij} = (2 \delta_{\sigma \rho} - 1)  
R^{\st}_{i \sigma} R^{\sb}_{j \rho}\, ,
\eeq
and $m_{\sg}$ is the gluino mass.
A gluon mass $\lambda$ is introduced to
regularize
the infrared divergences.
The UV divergences are regularized by dimensional reduction (DR)
\cite{Siegel,Jack} which preserves
supersymmetry at least at one--loop order.
We use the usual one--, two--, and three--point functions 
$A_0$, $B_0$, $B_1$, and
$C_0$ \cite{Denner}:
\bea
A_0(m^2) & = & \int\frac{d^Dq}{i\pi^2}\frac{1}{q^2-m^2}\,, \nonumber\\ 
\left[B_0, k^\mu B_1\right](k^2, m_1^2, m_2^2) & = &
\int\frac{d^Dq}{i\pi^2}\frac{[1, q^\mu]}
{(q^2-m_1^2)((q+k)^2-m_2^2)}\,, \nonumber\\
C_0(m_0^2, m_1^2, m_2^2)&=&
\int\frac{d^Dq}{i\pi^2}\frac{1}{(q^2-m_0^2)((q+k_{\sq_i})^2-m_1^2)
((q+k_{\sq_j})^2-m_2^2)} \nonumber\, .
\eea
In these integrals
$k_{\sq_i}$ and $- k_{\sq_j}$ are the external momenta of
$\sq_i$ and $\bar\sq_j$, respectively.\\ 
\\
The squark wave--function corrections $\delta G_{ijk}^{\sq\,(w)}$ 
can be expressed as
\beq \label{17}
\delta G_{ijk}^{\sq\,(w)}=-\frac{1}{2}\left[ \dot{\Pi}_{ii}^{\sq}(m_{\sq_i}^2)
+\dot{\Pi}_{jj}^{\sq}(m_{\sq_j}^2)\right] G_{ijk}^\sq
-\frac{\Pi_{ii'}^{\sq}(m_{\sq_i}^2)}{m_{\sq_i}^2-m_{\sq_{i'}}^2} G_{i'jk}^\sq
-\frac{\Pi_{j'j}^{\sq}(m_{\sq_j}^2)}{m_{\sq_j}^2-m_{\sq_{j'}}^2} G_{ij'k}^\sq
\, .
\eeq
Here and in the following $i\neq i'$ and $j\neq j'$.
$\Pi_{ij}^{\sq}(k^2)$ are the one--loop corrections to the two-point
functions of
$\bar{\sq}_i\sq_j$, which are obtained from the graphs of Fig.~1c.
$\dot{\Pi}(k^2)$
denotes the derivative with respect to $k^2$. The last two terms in (\ref{17})
represent the corrections due to squark mixing. Note that for
$H^+$~decay ($k = 4$) 
the subscripts $i$ and $i'$ are attached to $\st$ and
$j$ and $j'$ to $\sb$.
The explicit forms of the self--energies and their derivatives of the diagonal
parts are
\bea \label{27}
\Pi_{ij}^{\sq\,(g)}(k^2) & = & -\frac{\alpha_sC_F}{4\pi} \delta_{ij}
\left[(3 k^2 + m_{\sq_i}^2) B_0(k^2,\lambda^2 , m_{\sq_i}^2)+
2 k^2 B_1(k^2,\lambda^2 , m_{\sq_i}^2)\right]\, ,\\
\Pi_{ij}^{\sq\,(\sg)}(k^2) & = & -\frac{\alpha_sC_F}{\pi} 
\left[ \left( A_0(m_q^2) + k^2 B_1(k^2,m_\sg^2, m_q^2)\right) \delta_{ij}
\right.\nonumber\\
&& + \left( m_\sg^2  \delta_{ij} + (S^\sq)_{ij} m_q m_\sg \right)
B_0(k^2,m_\sg^2, m_q^2)\left. \right]\, ,
\eea
\beq \label{29}
\Pi_{ij}^{\sq\,(\sq)}(k^2) = \frac{\alpha_s C_F}{4\pi} \left(
\begin{array}{cc}
\cos^2 2 \theta_{\tilde q} A_0(m_{{\tilde q}_1}^2) +
 \sin^2 2 \theta_{\tilde q} A_0(m_{{\tilde q}_2}^2)&
 \frac{1}{2}\sin 4 \theta_{\tilde q}
 \left(A_0(m_{{\tilde q}_2}^2)- A_0(m_{{\tilde q}_1}^2)\right)\\
 \frac{1}{2}\sin 4 \theta_{\tilde q}
 \left(A_0(m_{{\tilde q}_2}^2)- A_0(m_{{\tilde q}_1}^2)\right) &
\sin^2 2 \theta_{\tilde q} A_0(m_{{\tilde q}_1}^2)
 + \cos^2 2 \theta_{\tilde q} A_0(m_{{\tilde q}_2}^2)
\end{array}\right)\, ,
\eeq
\bea
\dot{\Pi}_{ii}^{\sq\,(g)}(m_{\sq_i}^2)&=&
-\frac{\alpha_sC_F}{4\pi}\left[ 3B_0(m_{\sq_i}^2, 0, m_{\sq_i}^2)
+2B_1(m_{\sq_i}^2, 0, m_{\sq_i}^2) \right. \nonumber\\
&& +4m_{\sq_i}^2\dot{B}_0(m_{\sq_i}^2, \lambda^2, m_{\sq_i}^2)
+2m_{\sq_i}^2\dot{B}_1(m_{\sq_i}^2, 0, m_{\sq_i}^2)\,,\\
\dot{\Pi}_{ii}^{\sq\,(\sg)}(m_{\sq_i}^2)&=&
-\frac{\alpha_sC_F}{\pi}\left[ \right.
m_{\sg}^2\dot{B}_0(m_{\sq_i}^2, m_{\sg}^2, m_q^2) +B_1(m_{\sq_i}^2,
m_{\sg}^2, m_q^2) \nonumber\\
&&\left. + m_{\sq_i}^2\dot{B}_1(m_{\sq_i}^2, m_{\sg}^2, m_q^2)
+(-)^{i}\sin 2\theta_\sq m_qm_{\sg}\dot{B}_0(m_{\sq_i}^2, m_{\sg}^2,
m_q^2) \right]\, , \\
\dot{\Pi}_{ii}^{\sq\,(\sq)}(m_{\sq_i}^2)&=& 0 \, . 
\eea
\noindent Now we discuss the shifts $\delta G_{ij2}^{\sq\,(0)}$ in
eq.~(\ref{13}). From eqs.~(\ref{7}) and (\ref{8}) it follows:
\beq
\delta G_{ij2}^{\sq\,(0)} = \left(R^\sq\, \delta G_{LR}^\sq\, (R^{\sq})^T +
\delta R^\sq\, G_{LR}^\sq \,(R^{\sq})^T +
R^\sq\, G_{LR}^\sq\, \delta (R^{\sq})^T\right)_{ij}\, . 
\eeq 
Using eq.~(\ref{5}) one further gets:
\beq \label{34}
\delta G_{ij2}^{\sq\,(0)} = (R^\sq)_{ik} (\delta G_{LR}^\sq)_{kl} (R^\sq)_{jl}
- \left( (-1)^i G_{i'j2}^\sq + (-1)^j G_{ij'2}^\sq\right) 
\delta \theta_\sq \, .
\eeq
 From eqs.~(\ref{7}) and (\ref{8}) it directly follows: 
\bea
\delta G_{LR}^\st & = & -\frac{g}{2 m_W \si_\be} \left( \begin{array}{cc}
4 m_t \si_\al\delta m_t &
\delta(m_t A_t) \si_\al - \mu \co_\al \delta m_t  \\
\delta(m_t A_t) \si_\al - \mu \co_\al \delta m_t & 
4 m_t \si_\al\delta m_t 
\end{array}
\right)\, , \\
\delta G_{LR}^\sb & = & -\frac{g}{2 m_W \co_\be} \left( \begin{array}{cc}
4 m_b \co_\al\delta m_b &
\delta(m_b A_b) \co_\al - \mu \si_\al \delta m_b \\
\delta(m_b A_b) \co_\al - \mu \si_\al \delta m_b & 
4 m_b \co_\al\delta m_b 
\end{array}
\right)\, . 
\eea
To get the correction terms $\delta G_{ij1}^{\sq\,(0)}$ one makes the same
replacements as in eq.~(\ref{9}):
\beq
\delta G_{ij1}^{\sq\,(0)} = \left(
\delta G_{ij2}^{\sq\,(0)} \mbox{ with } \al \to
\al + \smallfrac \pi\right)\, .
\eeq
For the couplings to the $A^0$ boson, eqs.~(\ref{10}) and (\ref{11}) one gets
the correction terms
\beq \label{dGij30}
\delta G_{ij3}^{\sq\,(0)} = \frac{i g}{2 m_W}{\small \left( \begin{array}{cc}
0 & \delta(m_q A_q) \{\cot\be, \tan\be\} + \mu \delta m_q \\
-(\delta(m_q A_q) \{\cot\be, \tan\be\} + \mu \delta m_q ) & 0
\end{array}
\right)}\, ,
\eeq
where $\cot\beta$ ($\tan\beta$) has to be taken for $\sq = \st$ ($\sb$).\\
For the $H^+ \to \st \bar\sb$ ($k = 4$) we have
\beq \label{dGij40}
\delta G_{ij4}^{(0)} = \left(R^\st\, \delta G_{LR}\, (R^{\sb})^T\right)_{ij}-
 (-1)^i G_{i'j4} \delta \theta_\st -
 (-1)^j G_{ij'4} \delta \theta_\sb \, ,
\eeq
and
\beq
\begin{array}{l}
\delta G_{LR} = \\
 \frac{g}{\sqrt{2}m_W} \left(\begin{array}{cc}
2 m_b \delta m_b\tan\beta+2 m_t \delta m_t \cot\beta  &
\delta (m_b A_b)\tan\beta+ \delta m_b\mu \\
\delta (m_t A_t)\cot\beta+ \delta m_t\mu   &
2(\delta m_t m_b + m_t \delta m_b)/\sin 2\beta
 \end{array}\right)
\end{array} \, .
\eeq
\noindent
We now give the formulae for $\delta m_q$, $\delta(m_q A_q)$, and 
$\delta\theta_\sq$ in the on-shell scheme. 
These terms consist of three parts, denoted by the 
superscripts $g, \sg$, and $\sq$. Therefore we can write 
$\delta G_{ijk}^{\sq\,(0)} \equiv 
\delta G_{ijk}^{\sq\,(0,g)}+\delta G_{ijk}^{\sq\,(0,\sg)}
+\delta G_{ijk}^{\sq\,(0,\sq)}$.
Since the renormalized $m_q$ is taken to be the pole mass, one gets
\bea \label{37}
\delta m_q &=& \delta m_q^{(g)} + \delta m_q^{(\sg)} \, ,\nonumber\\
\delta m_q^{(g)} &=& -\frac{\alpha_sC_F}{2\pi}
m_q(B_0(m_q^2, 0, m_q^2)-B_1(m_q^2, 0, m_q^2))
\nonumber\, ,\mbox{ and}\\
\delta m_q^{(\sg)} &=& -\frac{\alpha_sC_F}{4\pi}
\left[\right.\sin 2\theta_\sq m_{\sg}(B_0(m_q^2, m_{\sg}^2, m_{\sq_1}^2)
-B_0(m_q^2, m_{\sg}^2, m_{\sq_2}^2))\nonumber\\
&&\mbox{\hphantom{1234578}}+m_q(B_1(m_q^2, m_{\sg}^2, m_{\sq_1}^2)
+B_1(m_q^2, m_{\sg}^2, m_{\sq_2}^2))\left.\right] \, ,
\eea
from the graphs of Fig.~1d. Note that $\delta m_q^{(\sq)} = 0$ 
because there is no corresponding Feynman graph.\\
The on--shell renormalization of the squark mixing angle $\theta_\sq$ 
and $m_q A_q$ is, however, not straightforward. Here we adopt the following 
procedure: We start from squark pole masses $m_{\sq_i}$ and 
the on--shell mixing angle $\theta_\sq$ which will be defined later. 
The other on--shell parameters ($M_{\tilde{Q},\tilde{U},\tilde{D}}$,
$A_q$) for squarks are then 
defined in terms of the above $m_{\sq_i}$ and $\theta_\sq$ 
by the tree-level relations eqs.~(\ref{1})--(\ref{5}). 
In this scheme $\delta(m_q A_q)$ takes the form 
\beq \label{dAqmq} 
\delta (m_q A_q) = \frac{1}{2}(\delta m_{\sq_1}^2 - \delta m_{\sq_2}^2)
\sin 2 \theta_\sq + (m_{\sq_1}^2 - m_{\sq_2}^2) \cos 2 \theta_\sq 
\delta\theta_\sq +
\delta m_q \mu \{ \cot \beta, \tan \beta\}\, ,
\eeq
where $\cot\be\, (\tan\be)$ is for $\sq = \st\, (\sb)$. 
Here one has 
$\delta m_{\sq_i}^2 = \mbox{Re}[\Pi_{ii}^{\sq\,(g)}(m_{\sq_i}^2) +
\Pi_{ii}^{\sq\,(\sg)}(m_{\sq_i}^2) +\Pi_{ii}^{\sq\,(\sq)}(m_{\sq_i}^2)]$
from eqs.~(\ref{27})-(\ref{29}).\\
Next we have to define the on--shell renormalized squark mixing angle 
$\theta_\sq$. We treated this problem in \cite{Eberl} in the case of
$e^+ e^- \to \sq_i \bar{\sq}_j$.
We fixed the counterterm of the mixing angle $\delta\theta_\sq$
such that it cancels the
off--diagonal part of the squark wave--function corrections to 
$e^+ e^- \to \sq_1 \bar{\sq}_2$. 
Here we use the same scheme and take 
$\delta\theta_\sq = \delta\theta_{\tilde q}^{(\tilde g)} +
\delta\theta_{\tilde q}^{(\tilde q)}$ from \cite{Eberl}:
\bea \label{42}
\delta\theta_{\tilde q}^{(\tilde g)} & = &
\frac{\alpha_s C_F}{4\pi}
\frac{m_{\sg} m_q}{I^{3L}_q (m_{\sq_1}^2 - m_{\sq_2}^2)}\left(
B_0(m_{\sq_2}^2,m_{\sg}^2,m_q^2) a_{11} -
B_0(m_{\sq_1}^2,m_{\sg}^2,m_q^2) a_{22}\right),\\ \label{43}
\delta\theta_{\tilde q}^{(\tilde q)} & = & 
\frac{\alpha_s C_F}{8\pi}\frac{\sin4\theta_\sq}{
m_{\sq_1}^2 - m_{\sq_2}^2} \left( A_0(m_{\sq_2}^2) - A_0(m_{\sq_1}^2)\right)\,,
\eea
with $a_{11} = 4(I^{3L}_q \cos^2\theta_\sq - s_W^2 e_q)$ and
$a_{22} = 4(I^{3L}_q \sin^2\theta_\sq - s_W^2 e_q)$. The counterterms 
$\delta G_{ijk}^{\sq\,(0)}$ are then completely fixed.\\

\indent 
In our calculation we need the on--shell parameters 
$(M_{\tilde{Q},\tilde{U},\tilde{D}},A_q)$. 
In a combined treatment of both
the stop and the sbottom sectors in the on--shell scheme
we have to pay special attention to
the parameter $M_{\tilde Q}$.
This is necessary for the calculation of
the decay width of $H^+ \to \st \bar\sb$ and
of the branching ratios
of Higgs decays.
At tree--level and in the \drbar scheme the parameter $M_{\tilde Q}$ 
in the stop and sbottom mass matrices must be equal because of 
SU(2)$_L$ symmetry. 
In the on--shell scheme, however, this is not the case. The shifts
from the \drbar parameters to the on--shell (i. e. physical) ones 
are different for the stop and sbottom sectors:
\beq
M^2_{\tilde Q}\Big|_\drbar = M^2_{\tilde Q}(\st)\Big|_{\mbox{os}} +
\delta M^2_{\tilde Q}(\st)\,,  \qquad
M^2_{\tilde Q}\Big|_\drbar = M^2_{\tilde Q}(\sb)\Big|_{\mbox{os}} +
\delta M^2_{\tilde Q}(\sb)\, ,
\eeq
with
\beq \label{eq48}
\delta M^2_{\tilde Q}(\sq) = \delta m_{\sq_1}^2 \cos^2 \theta_\sq 
+ \delta m_{\sq_2}^2 \sin^2 \theta_\sq -
(m_{\sq_1}^2 - m_{\sq_2}^2) \sin 2 \theta_\sq \delta \theta_\sq - 2 m_q
\delta m_q \, .
\eeq 
In this paper we take $M_{\tilde Q}(\st)\big|_{\mbox{os}}$ 
as the on--shell input parameter. This then leads to a 
shift of $M^2_{\tilde Q}$ in the sbottom sector:
\beq \label{49}
M^2_{\tilde Q}(\sb)\Big|_{\mbox{os}} = M^2_{\tilde Q}(\st)\Big|_{\mbox{os}}
+ \delta M^2_{\tilde Q}(\st) - \delta M^2_{\tilde Q}(\sb)\, .
\eeq
As all physical parameters are finite, the shift 
$\delta M^2_{\tilde Q}(\st) - \delta M^2_{\tilde Q}(\sb)$ has to 
be UV convergent. We have checked that this is indeed the case.\\

\noindent
The one--loop corrected decay width to ${\cal O}(\alpha_s)$
in the on--shell scheme is then given by
\beq \label{corrwidth}
\Gamma(H^k\rightarrow\sq_i\bar{\sq}_j)
=\frac{N_C\kappa}{16\pi m^3_{H^k}}[|G_{ijk}^\sq|^2 +2G_{ijk}^\sq{\rm
Re}(\delta G_{ijk}^{\sq\,(v)}
+\delta G_{ijk}^{\sq\,(w)}+\delta G_{ijk}^{\sq\,(0)})]. \label{20}
\eeq
We have checked the UV convergence of the amplitudes 
$G_{ijk}^{\sq\, \rm corr}$ of eq.~(\ref{13}) and hence also of
$\Gamma(H^k\rightarrow\sq_i\bar{\sq}_j)$.
The width of eq.~(\ref{corrwidth}) is still infrared divergent.\\

The infrared divergences in (\ref{20}) are cancelled by including the
${\cal O}(\alpha_s)$
contribution from real gluon emission from $\sq_i$ and $\bar{\sq}_j$ 
(see Fig.~1e).
The decay width of $H^k(p)\rightarrow\sq_i(k_1)+\bar{\sq}_j(k_2)+g(k_3)$ is
given by
\beq \label{width3}
\Gamma(H^k\rightarrow\sq_i\bar{\sq}_jg)
=\frac{\alpha_sC_FN_C|G_{ijk}^\sq|^2}{4\pi^2 m_{H^k}}
[(m_{H^k}^2-m_{\sq_i}^2-m_{\sq_j}^2)I_{12}
-m_{\sq_i}^2I_{11}-m_{\sq_j}^2I_{22}-I_1-I_2]. \label{23} \eeq
The functions $I_n$, and $I_{nm}$ are defined as \cite{Denner} \beq
I_{i_1\ldots i_n}=\frac{1}{\pi^2}
\int\frac{d^3k_1}{2E_1}\frac{d^3k_2}{2E_2}\frac{d^3k_3}{2E_3}
\delta^4(p-k_1-k_2-k_3)\frac{1}
{(2k_3k_{i_1}+\lambda^2)\ldots(2k_3k_{i_n}+\lambda^2)}. \eeq
The explicit forms of $I_{i_1\ldots i_n}$ are given in \cite{Denner}. 
In (\ref{23}),
$I_{11,22,12}$ are infrared divergent. We have checked that the infrared
divergences in
(\ref{23}) cancel those in (\ref{20}). In the numerical analysis we define
the corrected
decay width as
\beq \label{totwidth}
\Gamma^{\rm corr}(H^k\rightarrow\sq_i\bar{\sq}_j)\equiv
\Gamma(H^k\rightarrow\sq_i\bar{\sq}_j)+
\Gamma(H^k\rightarrow\sq_i\bar{\sq}_jg)\, .
\eeq


\section{Numerical results and conclusions}
We choose 
\{$m_{A^0}$, $m_{t,b}$, $M$, $\mu$, $\tan\beta$, $M_{\tilde{Q}}(\st)$, $A$\}
(with $M_{\tilde{Q}}(\st) \equiv M_{\tilde{Q}}(\st)\big|_{\mbox{os}}$) as the
basic input parameters of the MSSM, taking
$M=(\alpha_2/\alpha_s(m_\sg))m_{\sg}=(3/5\tan^2\theta_W)M'$,
$M_{\tilde{Q}}(\st) : M_{\tilde{U}} : M_{\tilde{D}} : M_{\tilde{L}} :
M_{\tilde{E}} = 1 : \frac 8 9 : \frac{10}{9} : 1 : 1$ and 
$A \equiv A_t = A_b = A_\tau$.
Here $M$ ($M'$) is
the SU(2) (U(1)) gaugino mass, $\alpha_2=g^2/4\pi$, and 
($M_{\tilde{L}, \tilde{E}}$,
$A_{\tau}$)
are the mass matrix parameters of the slepton sector \cite{Bartl1,Bartl2}.
We take $m_t = 175$~GeV, $m_b = 5$~GeV, $m_Z = 91.2$~GeV, $m_W = 80$~GeV,
$\sin^2\theta_W=0.23$, $\alpha_2 = 0.0337$,
and $\alpha_s = \alpha_s(m_{H^k})$ for the $H^k$ decay. We use
$\alpha_s(Q)=
12\pi/\{(33-2n_f)\ln(Q^2/\Lambda_{n_f}^2)\}$,
with $\alpha_s(m_Z)=0.12$, and the number of quark flavors $n_f=5(6)$ for
$m_b<Q\le m_t$ (for $Q>m_t$).\\

We define the QCD corrections as the difference between the ${\cal
O}(\alpha_s)$ corrected
width $\Gamma^{\rm
corr}(H^k\rightarrow\sq_i\bar{\sq}_j)$
of eq.~(\ref{totwidth}) (i. e. eqs.~(\ref{corrwidth}) plus 
(\ref{width3})) and the
tree--level width $\Gamma^{\rm tree}(H^k\rightarrow\sq_i\bar{\sq}_j)$
of eq.~(\ref{6}) with $M_{\tilde{Q}} = M_{\tilde{Q}}(\st)$ for both the
$\st$ and $\sb$ mass matrices. Note that $m_{\st_i} = m_{\st_i}^{\rm tree}$
but $m_{\sb_i} \ne m_{\sb_i}^{\rm tree}$ 
due to the shift in $M_{\tilde{Q}}(\sb)$ in eq.~(\ref{49}), where $m_{\sq_i}$ 
($m_{\sq_i}^{\rm tree}$) is the on--shell $\tilde{q}_i$--mass at one--loop
level (at tree--level). This shift is calculated by taking $\alpha_s$
at the scale $M_{\tilde{Q}}(\st)$ in eq.~(\ref{eq48}).\\

In order not to vary too many parameters, in the following we 
take the values of $M, \mu$, and $\tan\beta$ such 
that $m_{\tilde{\chi}^0_1} \simeq
70$~GeV as in \cite{Bartl2}, where $\tilde\chi^0_1$ is the lightest neutralino.
In
Fig.~2 we show the $m_{A^0}$ dependence of the tree--level and corrected
widths $\Gamma^{\rm tree}_{H^k}(\sq\bar{\sq}) \equiv \sum_{i,j = 1,2}
\Gamma^{\rm tree}(H^k \to \sq_i\bar{\sq}_j)$ and 
$\Gamma^{\rm corr}_{H^k}(\sq\bar{\sq}) \equiv \sum_{i,j = 1,2}
\Gamma^{\rm corr}(H^k \to \sq_i\bar{\sq}_j)$, and the tree-level
branching ratio
$B^{\rm tree}_{H^k}(\sq\bar{\sq}) \equiv$ $\sum_{i,j = 1,2}
B^{\rm tree}(H^k \to \sq_i\bar{\sq}_j)$ \cite{Bartl1,Bartl2} for
$(\tan\beta, M$~(GeV), $\mu$~(GeV), $M_{\tilde{Q}}(\st)$~(GeV), $A$~(GeV)) =
(2, 160, 300, 95, 300) (a, b, c), and (12, 140, --300, 150, --250) (d, e, f).
In these cases we have (in GeV units):
$(m_{\st_1}, m_{\st_2}, m_{\sb_1}, m_{\sb_2}, m_{\sb_1}^{\rm tree}, 
m_{\sb_2}^{\rm tree}, m_\sg, m_{\tilde{\chi}_1^+})$ = 
(106, 252, 105, 124, 99, 113, 465, 128)(a,b,c) and
(97, 297, 113, 215, 102, 210, 412, 133)(d,e,f).
Here $\tilde{\chi}_1^+$ is the lighter chargino. We see
that in these cases the $\st\bar{\sb}$ mode (the sum of the $\st\bar{\st}$ 
and $\sb\bar{\sb}$ modes) dominates the $H^+$ decay (the $H^0$ and $A^0$
decays) in a wide $m_{A^0}$ range at the tree--level, and that the QCD 
corrections to the $\st\bar{\sb}$ mode (the  $\st\bar{\st}$
and $\sb\bar{\sb}$ modes) are significant, but that as a whole they do not 
invalidate the dominance of the $\st\bar{\sb}$ mode (the sum of the
$\st\bar{\st}$ and $\sb\bar{\sb}$ modes). Our calculation
includes the leading Yukawa corrections to the Higgs sector as in
\cite{Bartl1,Bartl2}. Note that  $m_{H^+} \simeq m_{H^0} \simeq m_{A^0}$ in the
$m_{A^0}$ range shown here, and that the $A^0$ does not couple to
$\st_i\bar{\st}_i$ and $\sb_i\bar{\sb}_i$ $(i = 1,2)$. As for $h^0$ decay,
we have found that the decay $h^0 \to \st_1\bar{\st}_1$ is kinematically
allowed only in a very 
limited region of the MSSM parameter space \cite{Bartl2}.\\

In Table 1 we show the values of the tree--level branching ratios
$B^{\rm tree}(H^+ \to \st\bar{\sb})$ (a, d), 
$B^{\rm tree}(H^0 \to \st\bar{\st}, \sb\bar{\sb})$ (b, e),
$B^{\rm tree}(A^0 \to \st\bar{\st}, \sb\bar{\sb})$ (c, f), and the QCD
corrections $C(H^+ \to \st\bar{\sb})$ (a, d),
$C(H^0 \to \st\bar{\st}, \sb\bar{\sb})$ (b, e),
$C(A^0 \to \st\bar{\st}, \sb\bar{\sb})$ (c, f) for typical values of
$M_{\tilde{Q}}(\st)$ and $A$, for ($m_{A^0}$~(GeV), $\tan\beta$, $\mu$~(GeV),
$M$~(GeV))
= (400, 2, 300, 160) (a), (450, 2, 300, 160) (b), 
(500, 2, 300, 160) (c), (400, 12, --300, 140) (d),
(450, 12, --300, 140) (e), and (500, 12, --300, 140) (f). Here 
$B^{\rm tree}(H^+ \to \st\bar{\sb}) \equiv B^{\rm tree}_{H^+} (\st\bar{\sb})$,
$B^{\rm tree}(H^k \to \st\bar{\st}, \sb\bar{\sb}) \equiv
\sum_{\sq = \st, \sb} B^{\rm tree}_{H^k} (\sq\bar{\sq})\ (k = 2,3)$,
$C(H^+ \to \st\bar{\sb}) \equiv (\Gamma_{H^+}^{\rm corr} (\st\bar{\sb}) -
\Gamma_{H^+}^{\rm tree} (\st\bar{\sb}))/\Gamma_{H^+}^{\rm tree} (\st\bar{\sb})$,
and $C(H^k \to \st\bar{\st}, \sb\bar{\sb}) \equiv$ $( \sum_{\sq = \st, \sb}
\Gamma_{H^k}^{\rm corr} (\sq\bar{\sq}) -
\sum_{\sq = \st, \sb} \Gamma_{H^k}^{\rm tree} (\sq\bar{\sq}))/
\sum_{\sq = \st, \sb} \Gamma_{H^k}^{\rm tree} (\sq\bar{\sq})\ (k = 2,3)$.
We see again that the QCD corrections are significant, but that 
the $\sq\bar{\sq}$ modes dominate the $H^+$, $H^0$, and $A^0$ decays in
a wide region also when the QCD corrections are included. We have found
that our results are rather insensitive to the assumptions on the ratios of 
$M_{\tilde U, \tilde D, \tilde L, \tilde E}/M_{\tilde Q}(\st)$ and
$A_{b, \tau}/A_t$.\\ 
Here we note that for large $\tan\beta$ (and large $|\mu|$)
we often get negative corrected widths for some of the $\sb$--involved modes
(i.~e. the $\sb_i\bar\sb_j$ and $\st_i\bar\sb_j$ modes) depending on the values
of the other input parameters. This is mainly due to a large value of
the third term of $\delta (A_b m_b)$ of eq.~(\ref{dAqmq}) for large $\tan\beta$
which leads to large values of the shifts $\delta G_{ijk}^{\sb\, (0)}$, 
$(k = 2, 3)$ and $\delta G_{ij4}^{(0)}$ of eqs.~(\ref{34}), (\ref{dGij30}),
and (\ref{dGij40}), which then can result in negative corrected widths
for some of the $\sb$--involved modes. Here note that the shifts 
$\delta G_{ijk}^{\sb\, (0)}$ and $\delta G_{ij4}^{(0)}$ are roughly 
proportional to 
$\delta (A_b m_b) \tan\beta \sim \delta m_b\, \mu \tan^2\beta$.\\

In conclusion, we have calculated the ${\cal O}(\alpha_s)$ QCD corrections
to the decay widths of $H^+ \to \st\bar{\sb}$ and $H^0, A^0 \to \st\bar{\st},
\sb\bar{\sb}$ in the on--shell scheme, including all quark mass terms and
$\sq_L$--$\sq_R$ mixing. We find that the QCD corrections are significant,
but that they do not invalidate our previous conclusions at tree--level
about the dominance of the $\st\bar{\sb}$, and $\st\bar{\st}, \sb\bar{\sb}$
modes in a wide MSSM parameter region.

\section*{Acknowledgements}
The work of A.~B., H.~E., and W.~M. was supported by the ``Fonds zur
F\"orderung der
wissenschaftlichen Forschung'' of Austria, project no. P10843-PHY.

\vspace{6mm}
\section*{Figure Captions}
\renewcommand{\labelenumi}{Fig. \arabic{enumi}} \begin{enumerate}

\vspace{6mm}
\item
All diagrams relevant for the calculation of the ${\cal O}(\alpha_s)$
QCD corrections to the width of $H^k\rightarrow\sq_i\bar{\sq}_j$ in
the MSSM.

\vspace{6mm}
\item
The $m_{A^0}$ dependence of $\Gamma_{H^k}^{\rm tree} (\sq\bar{\sq})$ 
(dashed line), $\Gamma_{H^k}^{\rm corr} (\sq\bar{\sq})$ (solid line),
and $B_{H^k}^{\rm tree} (\sq\bar{\sq})$ (short--dashed line) for 
$(\tan\beta, M$~(GeV), $\mu$~(GeV), $M_{\tilde{Q}}(\st)$~(GeV), $A$~(GeV)) =
(2, 160, 300, 95, 300) (a, b, c), and (12, 140, --300, 150, --250)(d, e, f).

\end{enumerate}

\vspace{6mm}

\section*{Table Caption}
\renewcommand{\labelenumi}{Table \arabic{enumi}} \begin{enumerate}
\vspace{6mm}

\item
$B^{\rm tree}$ and $C$ for typical values of $M_{\tilde{Q}}(\st)$ and $A$, for
various values of ($m_{A^0}, \tan\beta,\mu , M$). See the text for details.
The requirement $m_{\st_1, \sb_1, \tilde l^{^{-}}} > m_{\tilde{\chi}_1^0}$ ($\simeq 70$~GeV)
is satisfied for these parameter values. 

\end{enumerate}

\clearpage

\begin{center}
\begin{picture}(412,556)
\put(0,0){\mbox{\psfig{file=H0graphs.eps}}}
\setlength{\unitlength}{1mm}
\put(74,156){\makebox(0,0)[t]{\large{\bf{Fig.~1a}}}}
\put(55,180){\makebox(0,0)[bl]{$H^k$}}
\put(95,197){\makebox(0,0)[l]{$\sq_i$}}
\put(95,161){\makebox(0,0)[l]{$\bar{\sq}_j$}}

\put(74,101){\makebox(0,0)[t]{\large{\bf{Fig.~1b}}}}
\put(3,126){\makebox(0,0)[bl]{$H^k$}}
\put(43.5,143){\makebox(0,0)[l]{$\sq_i$}}
\put(29,132){\makebox(0,0)[b]{$\sq_i$}}
\put(29,118){\makebox(0,0)[t]{$\bar{\sq}_j$}}
\put(36,125){\makebox(0,0)[l]{$g$}}
\put(43.5,107){\makebox(0,0)[l]{$\bar{\sq}_j$}}

\put(55,126){\makebox(0,0)[bl]{$H^k$}}
\put(81,132){\makebox(0,0)[b]{$q$}}
\put(81,117){\makebox(0,0)[t]{$\bar q$}}
\put(86,125){\makebox(0,0)[l]{$\sg$}}
\put(95,143){\makebox(0,0)[l]{$\sq_i$}}
\put(95,107){\makebox(0,0)[l]{$\bar{\sq}_j$}}

\put(106.5,126){\makebox(0,0)[bl]{$H^k$}}
\put(122.5,132.5){\makebox(0,0)[b]{$\sq$}}
\put(122.5,116.5){\makebox(0,0)[t]{$\bar{\sq}$}}
\put(147,143){\makebox(0,0)[l]{$\sq_i$}}
\put(147,107){\makebox(0,0)[l]{$\bar{\sq}_j$}}

\put(73,56){\makebox(0,0)[t]{\large{\bf{Fig.~1c}}}}
\put(2,66.5){\makebox(0,0)[tl]{$\sq_i$}}
\put(14.5,66.5){\makebox(0,0)[t]{$\sq_i$}}
\put(14.5,77.5){\makebox(0,0)[b]{$g$}}
\put(27,66.5){\makebox(0,0)[tr]{$\sq_i$}}

\put(41,66.5){\makebox(0,0)[tl]{$\sq_i$}}
\put(53.5,83.5){\makebox(0,0)[b]{$g$}}
\put(66,66.5){\makebox(0,0)[tr]{$\sq_i$}}

\put(80,66.5){\makebox(0,0)[tl]{$\sq_j$}}
\put(92.5,66.5){\makebox(0,0)[t]{$q$}}
\put(92.5,76.5){\makebox(0,0)[b]{$\sg$}}
\put(105,66.5){\makebox(0,0)[tr]{$\sq_i$}}

\put(119,66.5){\makebox(0,0)[tl]{$\sq_j$}}
\put(131.5,82.5){\makebox(0,0)[b]{$\sq$}}
\put(144,66.5){\makebox(0,0)[tr]{$\sq_i$}}

\put(21,-7){\makebox(0,0)[t]{\large{\bf{Fig.~1d}}}}
\put(8.5,22){\makebox(0,0)[tl]{$q$}}
\put(21,22){\makebox(0,0)[t]{$q$}}
\put(21,32.2){\makebox(0,0)[b]{$g$}}
\put(33.5,22){\makebox(0,0)[tr]{$q$}}

\put(8.5,3.1){\makebox(0,0)[tl]{$q$}}
\put(21,3.1){\makebox(0,0)[t]{$\sq$}}
\put(21,12.3){\makebox(0,0)[b]{$\sg$}}
\put(33.5,3.1){\makebox(0,0)[tr]{$q$}}

\put(99.8,-7){\makebox(0,0)[t]{\large{\bf{Fig.~1e}}}}
\put(54.8,18.7){\makebox(0,0)[bl]{$H^k$}}
\put(94.8,35.7){\makebox(0,0)[l]{$\sq_i$}}
\put(94.8,26){\makebox(0,0)[l]{$g$}}
\put(94.8,-0.3){\makebox(0,0)[l]{$\bar{\sq}_j$}}

\put(105.8,18.7){\makebox(0,0)[bl]{$H^k$}}
\put(146.8,35.7){\makebox(0,0)[l]{$\sq_i$}}
\put(146.8,9){\makebox(0,0)[l]{$g$}}
\put(146.8,-0.3){\makebox(0,0)[l]{$\bar{\sq}_j$}}
\end{picture}
\end{center}

\hspace{-2.8cm}\mbox{\psfig{file=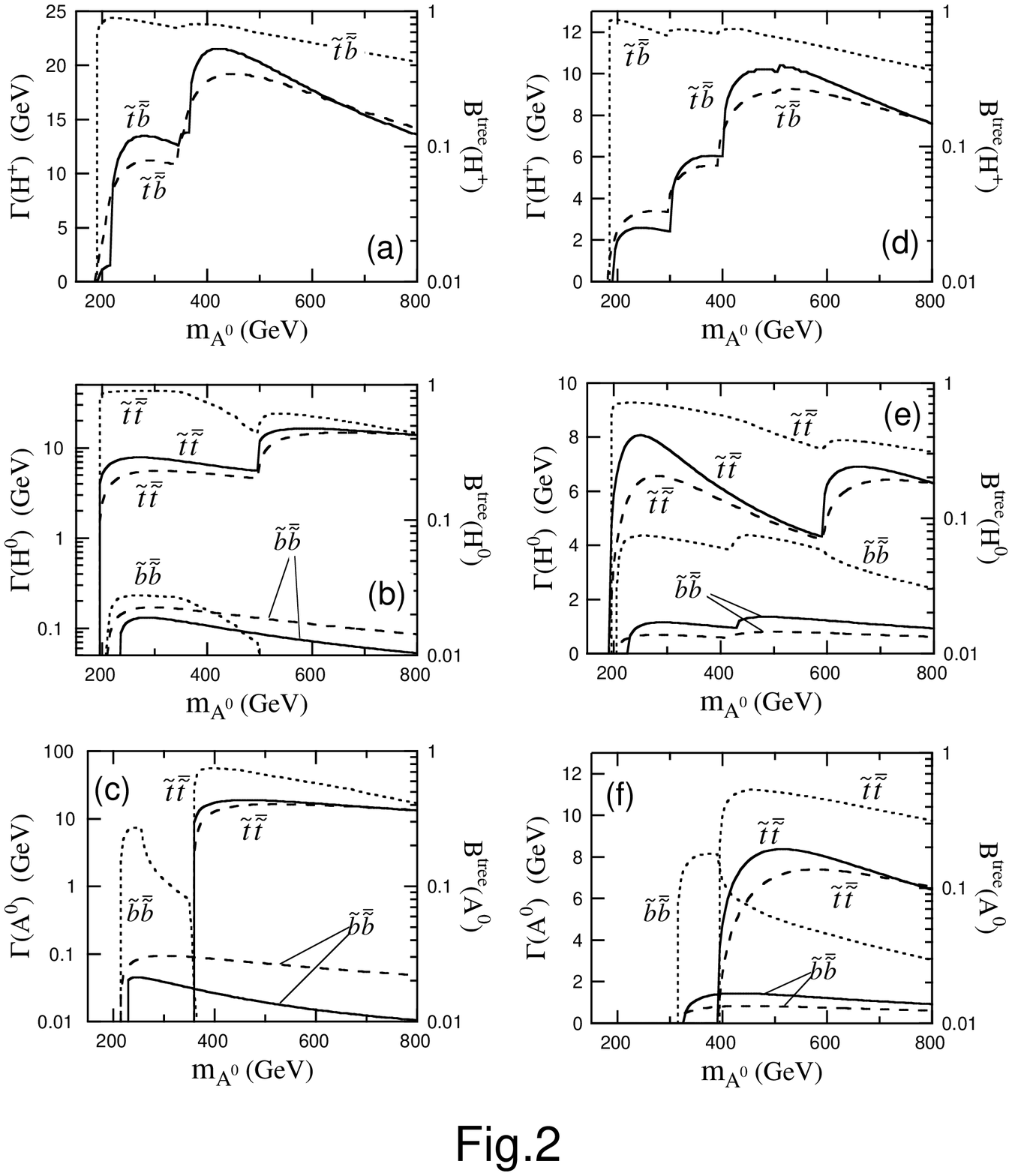}}

\clearpage
~~~\\
\vfill

\begin{center}
{{\large \bf Table 1}}\\
\end{center}
\hspace{-1cm}
{\small 
\begin{tabular}{|rccrr|rccrr|}
\hline
 & ${M_{\widetilde{Q}}({\widetilde{t}})}$(GeV) & A(GeV) & B$^{tree}$ & C \qquad & 
 & ${M_{\widetilde{Q}}({\widetilde{t}})}$(GeV) & A(GeV) & B$^{tree}$ & C \qquad \\
\hline
(a) & $80$ & $   0$ & $0.704$ & $-0.002$ & (d) & $140$ & $-250$ &  $0.781$ & $ 0.158$ \\
    & $80$ & $ 250$ & $0.803$ & $ 0.125$ &     & $140$ & $   0$ &  $0.735$ & $ 0.121$ \\
    &$120$ & $ 150$ & $0.732$ & $ 0.148$ &     & $140$ & $ 200$ &  $0.756$ & $ 0.091$ \\
    &$120$ & $ 350$ & $0.751$ & $-0.052$ &     & $180$ & $-300$ &  $0.680$ & $ 0.114$ \\
    &$140$ & $   0$ & $0.609$ & $ 0.002$ &     & $180$ & $   0$ &  $0.624$ & $ 0.196$ \\
    &$160$ & $ 350$ & $0.678$ & $ 0.185$ &     & $180$ & $ 250$ &  $0.646$ & $ 0.159$ \\
    &$180$ & $-150$ & $0.506$ & $-0.103$ &     & $220$ & $-400$ &  $0.669$ & $ 0.103$ \\
    &$240$ & $ 550$ & $0.706$ & $ 0.241$ &     & $220$ & $ 350$ &  $0.600$ & $ 0.135$ \\
\hline
(b) & $80$ & $   0$ & $0.818$ & $-0.072$ & (e) & $140$ & $-250$ &  $0.595$ & $ 0.201$ \\
    & $140$ & $ 100$ & $0.752$ & $ 0.164$ &     & $180$ & $200$ &  $0.560$ & 
$0.116$ \\
    &$200$ & $-150$ & $0.746$ & $-0.132$ &     & $200$ & $ 250$ &  $0.564$ & $ 0.155$ \\ 
    &$200$ & $   0$ & $0.706$ & $ 0.131$ &     & $240$ & $ -400$ &  $0.569$ & $ 0.018$ \\
    &$200$ & $ 400$ & $0.644$ & $ 0.220$ &     & $260$ & $ 400$ &  $0.536$ & $ 0.092$ \\
    &$260$ & $-300$ & $0.649$ & $-0.247$ &     & $300$ & $-600$ &  $0.611$ & $ 0.011$ \\
    &$260$ & $ 600$ & $0.777$ & $ 0.153$ &     & $340$ & $ 650$ &  $0.497$ & $ 0.075$ \\
    &$360$ & $ 900$ & $0.865$ & $ 0.171$ &     & $400$ & $-950$ &  $0.649$ & $ 0.005$ \\
\hline
(c) & $80$ & $ 200$ & $0.617$ & $ 0.135$ & (f) & $140$ & $-250$ &  $0.591$ & $ 0.270$ \\
    & $80$ & $ 300$ & $0.680$ & $ 0.118$ &     & $140$ & $ -50$ &  $0.503$ & $ 0.350$ \\
    & $100$ & $ 300$ & $0.668$ & $ 0.139$ &     & $160$ & $   0$ &  $0.445$ & $ 0.389$ \\ 
    &$120$ & $ 250$ & $0.616$ & $ 0.185$ &     & $170$ & $ 150$ &  $0.440$ & $ 0.318$ \\
    &$140$ & $ 200$ & $0.550$ & $ 0.244$ &     & $180$ & $-200$ &  $0.464$ & $ 0.302$ \\
    &$140$ & $ 350$ & $0.661$ & $ 0.186$ &     & $180$ & $ 200$ &  $0.433$ & $ 0.345$ \\
    &$180$ & $ 300$ & $0.505$ & $ 0.451$ &     & $200$ & $-350$ &  $0.531$ & $ 0.071$ \\
    &$180$ & $ 450$ & $0.672$ & $ 0.221$ &     & $200$ & $ 300$ &  $0.434$ & $ 0.402$ \\
\hline
\end{tabular}
\label{table1}
}

\vfill

\end{document}